\documentclass[twocolumn,aps,prd]{revtex4}
 \usepackage{graphicx,amssymb}
 \usepackage[utf8]{inputenc} 
 
\usepackage[urlcolor=blue,colorlinks,breaklinks=true]{hyperref}

\PassOptionsToPackage{obeyspaces,spaces,hyphens}{url}
 
 \usepackage{amsmath}
 
 \usepackage{txfonts,upgreek, bm}
 
\begin{document}
 	\def\half{{1\over2}}
 	\def\shalf{\textstyle{{1\over2}}}
 	
 	\newcommand\lsim{\mathrel{\rlap{\lower4pt\hbox{\hskip1pt$\sim$}}
 			\raise1pt\hbox{$<$}}}
 	\newcommand\gsim{\mathrel{\rlap{\lower4pt\hbox{\hskip1pt$\sim$}}
 			\raise1pt\hbox{$>$}}}

\newcommand{\be}{\begin{equation}}
\newcommand{\ee}{\end{equation}}
\newcommand{\bq}{\begin{eqnarray}}
\newcommand{\eq}{\end{eqnarray}}

\title{Particles and their fluids in nontrivial matter extensions to general relativity}
 	 	
\author{P.P. Avelino}
\email[Electronic address: ]{pedro.avelino@astro.up.pt}
\affiliation{Departamento de F\'{\i}sica e Astronomia, Faculdade de Ci\^encias, Universidade do Porto, Rua do Campo Alegre 687, PT4169-007 Porto, Portugal}
\affiliation{Instituto de Astrof\'{\i}sica e Ci\^encias do Espa{\c c}o, Universidade do Porto, CAUP, Rua das Estrelas, PT4150-762 Porto, Portugal}
\affiliation{Université Côte d’Azur, Observatoire de la Côte d’Azur, CNRS, Laboratoire Lagrange, France}

\date{\today}
\begin{abstract}
	
According to the standard von Laue condition, the volume-averaged pressure inside particles of fixed mass and structure vanishes in the Minkowski limit of general relativity. Here we show that this condition is in general not  fulfilled in the context of $f(R,T)$ gravity, or of other theories of gravity in which the linear momentum is not conserved in this limit (here, $R$ and $T$ represent the Ricci scalar and the trace of the energy-momentum tensor, respectively). We derive a generalized von Laue condition valid for the $\mathcal R(R) + \mathcal F(T)$ subclass of $f(R,T)$ theories of gravity and discuss its cosmological implications. In particular, we show that the standard radiation and matter era evolution of the universe is recovered in the context $R + \mathcal F(T)$ gravity independently of the specific properties of the function $\mathcal F(T)$.  We also find that dust --- a perfect fluid whose particles are at rest in the fluid's proper frame --- cannot in general be described as pressureless in the context of these theories. We further discuss the implications of our findings for the form of the on-shell Lagrangian of an ideal gas.

\end{abstract}

\maketitle
 	
\section{Introduction}
\label{sec:intr}

In the Minkowski limit of general relativity, the equations of motion for the matter fields may be obtained from the matter Lagrangian using a standard variational principle. The existence of stable compact objects in this limit requires that the volume-averaged pressure inside those objects vanishes (see \cite{doi:10.1002/andp.19113400808} for the original derivation of this condition by  von Laue, as well as \cite{Avelino:2018qgt,Avelino:2023rac} for alternative derivations). The standard von Laue condition has been shown to apply to particles, here defined as stable compact objects of fixed proper mass and structure with negligible self-induced metric perturbations, but also to the transverse pressure of defects of co-dimension $D < N$ in $N + 1$-dimensional space-times \cite{Avelino:2018qgt}. In fact, it is also valid in the context of modified gravity as long as the equations of motion determining the structure of particles or defects are the same as those obtained in the Minkowski limit of general relativity --- here defined as the limit where the  self-induced gravitational field is too weak to have a significant impact on the structure of particles or defects. The average pressure inside static objects can deviate from zero in general relativity (see, for example, \cite{Batool:2024axw}),  but only if gravity plays a significant role in the object's structure.

In general relativity and other theories of gravity perfect fluids are often employed to describe the material content of the Universe without an explicit reference to individual particles nor to the form of the Lagrangians which describe their dynamics \cite{Schutz:1970my, Schutz:1977df,  Brown:1992kc,Faraoni:2009rk,Capozziello:2018ddp,Ferreira:2020fma}.  However, in modified gravity (see \cite{Capozziello:2011et,Olmo:2011uz,Clifton:2011jh,Berti:2015itd,Nojiri:2017ncd,Odintsov:2023weg} for recent reviews) the knowledge of the on-shell matter Lagrangian might be crucial for an accurate computation of the dynamics of the gravitational and matter fields, in particular if the matter fields are non-minimally coupled to the geometry \cite{Nesseris:2008mq,Harko:2010mv,Harko:2011kv,Katirci:2013okf,Haghani:2013oma,Ludwig:2015hta,Harko:2018gxr,Bahamonde:2017ifa, Barrientos:2018cnx,Minazzoli:2018xjy,Fox:2018gop,Asimakis:2022jel,Goncalves:2023umv}. Although there is no universal on-shell Lagrangian of a perfect fluid \cite{Ferreira:2020fma}, it has been shown that the standard von Laue condition implies that the on-shell Lagrangian of an ideal gas --- or, in fact, of any fluid that can be approximated as a collection of moving localized particles of fixed proper mass and structure ---  is equal to the trace of the fluid energy-momentum tensor, assuming again that the equations of motion determining the particle structure are the same as those in general relativity in the Minkowski limit \cite{Avelino:2018qgt,Avelino:2018rsb,Avelino:2022eqm}. Notice that a significant fraction of the energy content of the Universe, including dark matter, baryons and photons (but not dark energy), may be described using an ideal gas approximation.

In $f(R,T)$ gravity \cite{Harko:2011kv} not only the on-shell matter Lagrangian appears explicitly in the equations of motion of the gravitational and matter fields, but also their dynamics depends on the first variation of the trace of the energy-momentum tensor with respect to the metric \cite{Haghani:2023uad,Avelino:2024rub}. Generally this implies that the equations of motion of the matter fields are modified with respect to general relativity, with the energy-momentum tensor not being covariantly conserved even when considering the Minkowski limit of $f(R,T)$ gravity. This can affect the particles' structure, potentially leading to a breakdown of the standard von Laue condition and affecting the form of the on-shell Lagrangian of the corresponding fluids. In this work, we shall investigate such breakdown in the context of theories of gravity in which energy and momentum conservation does not generally hold (even in the Minkowski limit), considering $f(R,T)$ gravity as a representative example. 

We shall examine in particular the $R + \mathcal F(T)$ subclass of $f(R,T)$ theories since it will turn out to be a well controlled family of gravity models due to its equivalence to general relativity with a modified energy-momentum tensor. We will derive a generalized von Laue condition valid in $\mathcal R (R) + \mathcal F(T)$ gravity (here, $\mathcal R (R)$ is an arbitrary function of $R$) and discuss its implications for the properties of cosmic fluids composed of relativistic or non-relativistic particles as well as the corresponding impact on the evolution of the Universe. In particular we will critically assess recent claims \cite{Haghani:2023uad}, obtained implicitly assuming the validity of the standard von Laue condition, that the matter era evolution of the universe in $R + \mathcal F(T)$ gravity should depend on the specific properties of the function $\mathcal F(T)$.  We also discuss the implications of our findings for the form of the on-shell Lagrangian of an ideal gas in the context of these theories.

Throughout this work, we will employ units where $c=16\pi G = \hbar=1$ with $c$, $\hbar$, and $G$ being, respectively, the speed of light in vacuum, the reduced Planck constant, and Newton's gravitational constant. We also adopt the metric signature $(-,+,+,+)$. The Greek indices and the Latin indices $i$ and $j$ take the values $0,1,2,3$ and $1,2,3$ (or, equivalently, $0,x,y,z$ and $x,y,z$), respectively. The Einstein summation convention will be used when a Greek or Latin index appears twice in a single term, once in an upper (superscript) and once in a lower (subscript) position.

\section{$f(R,T)$ gravity}

$f(R,T)$ gravity is defined by the action
\be
\label{eq:action}
S=\int d^4 x \sqrt{-g} \left[ f(R,T) + \mathcal{L}_{\rm m}\right]\,,
\ee
where $g$ is the determinant of the metric $g_{\mu\nu}$, $\mathcal{L}_{\rm m}$ is the Lagrangian of the matter fields, and $f(R,T)$ is a generic function of the Ricci scalar $R\equiv R^{\mu \nu} g_{\mu \nu}$ and of the trace of the energy-momentum tensor $T \equiv T^{\mu \nu} g_{\mu \nu}$. The corresponding equations of motion for the gravitational field are given by
 \cite{Harko:2011kv}
\be
2(R_{\mu\nu} - \Delta_{\mu\nu})f_{,R} - g_{\mu\nu}R={\mathcal T}_{\mu\nu}\, ,
\ee
where a comma denotes a partial derivative, $R_{\mu\nu}$ is the Ricci tensor, $\Delta_{\mu \nu} \equiv \nabla_\mu \nabla_\nu - g_{\mu \nu} \Box$, $\Box \equiv \nabla^\mu \nabla_\mu$,
\bq
{\mathcal T}_{\mu\nu}&=&  T_{\mu\nu}+(f-R)g_{\mu\nu}-2f_{,T}(T_{\mu\nu}+{\mathbb  T}_{\mu\nu}) \,, \label{MEMT0}\\
T_{\mu\nu} &=& -{2\over \sqrt{-g}}{\delta(\sqrt{-g}\mathcal{L}_{\rm m})\over \delta g^{\mu\nu}}= g_{\mu \nu} \mathcal{L}_{\rm m} -2 \frac{\delta\mathcal{L}_{\rm m}}{ \delta g^{\mu\nu}}\,, \label{EMTC}\\
{\mathbb{T}}_{\mu \nu} &=& g^{\alpha \beta}\frac{\delta T_{\alpha \beta} }{\delta g^{\mu \nu} }=\frac{\delta T}{\delta g^{\mu \nu}}-T_{\mu\nu}\nonumber\\
&=&-2T_{\mu \nu} +g_{\mu \nu} \mathcal{L}_{\rm m} - 2 g^{\alpha \beta} \frac{\delta^2 \mathcal{L}_{\rm m}}{\delta g^{\mu \nu} \delta g^{\alpha \beta} } \,,
\eq
and
\be
\nabla^\mu {\mathcal T}_{\mu\nu} =  2 (f_{,R}-1) \nabla^\mu  R_{\mu \nu}\,.
\ee
The condition $f_{,R}=1$, or equivalently $f(R,T)=R+\mathcal F(T)$ (here $\mathcal F(T)$ is a generic function of $T)$, is a necessary and sufficient condition for the covariant conservation of ${\mathcal T}_{\mu\nu}$. However, more generally, if $f(R,T)=\mathcal R (R)+\mathcal F(T)$, where $\mathcal R$ is an arbitrary function of $\mathcal R$, then
\be
{\mathfrak T}_{\mu\nu}=  T_{\mu\nu}+\mathcal F g_{\mu\nu}-2{\mathcal F}_{,T}(T_{\mu\nu}+{\mathbb  T}_{\mu\nu}) \label{TGR}
\ee
is covariantly conserved, so that
\be
\nabla^\mu {\mathfrak T}_{\mu\nu}=0\,. \label{conserv}
\ee 

If  $f(R,T)=R+\mathcal F(T)$ then
\be
G_{\mu\nu}=R_{\mu\nu} -\frac12 g_{\mu\nu}R=\frac12 {\mathfrak T}_{\mu\nu} \label {GR}\,.
\ee
Equation \eqref{GR} is just a modified version of the Einstein equations where ${\mathfrak{T}}_{\mu \nu}$ plays the role of $T_{\mu \nu}$.  $R+\mathcal F(T)$ gravity is, in fact, equivalent to general relativity with the modified matter Lagrangian  \cite{Fisher:2019ekh,Akarsu:2023lre,Avelino:2024rub}
\be
\mathfrak L_{\rm m}=\mathcal{L}_{\rm m}+\mathcal F\,.
\ee
This equivalence is particularly useful in this context, because it will allow us to investigate features of more general $f(R,T)$ gravity theories with a  covariantly conserved energy-momentum tensor considering a familiar set of models. Notice that, in this class of gravity models and, more generally, in $\mathcal R (R)+\mathcal F(T)$ gravity
\be
{\mathfrak T_{\mu\nu}} = -{2\over \sqrt{-g}}{\delta(\sqrt{-g}\mathfrak{L}_{\rm m})\over \delta g^{\mu\nu}}\,. \label{TN}
\ee
For this reason, we shall refer to ${\mathfrak T_{\mu\nu}} $ as the (covariantly conserved) modified energy-momentum tensor. Notice that ${\mathfrak T}_{\mu\nu}={\mathcal T}_{\mu\nu}$ in $R+\mathcal F(T)$ gravity, but that is not generally the case in $f(R,T)$ gravity. For example, in $\mathcal R (R)+\mathcal F(T)$ gravity the energy-momentum tensor defined in Eq. \eqref{TN} is still given by Eq. \eqref{TGR}, but in general one has ${\mathfrak T}_{\mu\nu} \neq {\mathcal T}_{\mu\nu}$. In fact, in $\mathcal R (R)+\mathcal F(T)$ gravity the modified energy-momentum tensor ${\mathfrak T}_{\mu\nu}$ will be covariantly conserved ($\nabla^\mu {\mathfrak T}_{\mu\nu}=0$), but ${\mathcal T}_{\mu\nu}$ will generally not ($\nabla^\mu {\mathcal T}_{\mu\nu} \neq 0$).

\section{The von Laue condition \label{SVLC}}

Here we will follow von Laue's reasoning, but applied to the covariantly conserved tensor ${\mathfrak T_{\mu\nu}}$ (in $\mathcal R (R)+\mathcal F(T)$ gravity) rather than to the energy-momentum tensor $T_{\mu\nu}$  --- notice that the latter is in general not covariantly conserved in $f(R,T)$ gravity, or even in its $R+\mathcal F(T)$ subclass (this is so also in the Minkowski limit). 

Consider a static localized particle with fixed mass and structure in a locally Minkowski spacetime
 where the line element is locally given by 
\be
ds^2=-dt^2+dx^2+dy^2+dz^2\,,
\ee 
where $x$, $y$ and $z$ are cartesian coordinates. Eq. \eqref{conserv} then implies that
\be
\partial^i {\mathfrak T}_{i j}=0\,.
\ee
On the other hand
\be
\int_{\mathcal V} \partial^i {\mathfrak T}_{i j} d^3 r = \oint_{\mathcal S}  {\mathfrak T}_{i j} n^i dS = 0 \,,
\ee
where ${\mathcal V}$ represents the integration volume, ${\mathcal S}$ is the surface bounding that volume, and $n^i$ are the components of the unit vector normal to the surface at each point (pointing outwards). Evaluating the surface integral on a plane with constant $x$ and closing it at infinity, one obtains that
\be
\int_{-\infty}^{+\infty} \int_{-\infty}^{+\infty}{\mathfrak T}_{x j} dy dz = 0 \,.
\ee
Finally, integrating over $x$, one finds that 
\be
\int_{-\infty}^{+\infty} \int_{-\infty}^{+\infty} \int_{-\infty}^{+\infty} {\mathfrak T}_{x j} dx dy dz= 0 \,,
\ee
the same reasoning applying if the cartesian coordinate $x$ is replaced by $y$ or $z$. Therefore
\be
\int {\mathfrak T}_{i j} d^3 r= 0 \,,
\ee
where the volume integral is over all space.
Defining the proper modified pressure as
\be
\mathcal P \equiv ({\mathfrak T}_{x x} + {\mathfrak T}_{y y} + {\mathfrak T}_{z z})/3\,,
\ee 
one finds the generalized von Laue condition
\be
\int  \mathcal P d^3 r= 0 \label{MVL}\,,
\ee
or, equivalently,
\be
\int  \left(p+ \mathcal F -2{\mathcal F}_{,T}(p+{\mathbb  P})\right)  d^3 r= 0 \,,
\ee
where $p\equiv (T_{x x} +T_{y y} + T_{z z})/3$ is the proper pressure and $\mathbb P \equiv ({\mathbb T}_{x x} + {\mathbb T}_{y y} + {\mathbb T}_{z z})/3$.

Notice that the standard von Laue condition ($\int p d^3 r = 0$) is not expected to apply in $f(R,T)$ gravity (nor even in its $\mathcal R (R)+\mathcal F(T)$ or $R+\mathcal F(T)$ subclasses)  since the energy-momentum tensor is generally not covariantly conserved --- a concrete illustrative example in $1+1$ dimensions will be considered in the forthcoming section. Pressure is a source of gravity and, therefore, this result can have profound implications, notably for the evolution of the Universe as a whole.

\subsection{Implications for the form of the on-shell Lagrangian}

The conservation of the modified proper mass of a particle defined by 
\be
\mathfrak m \equiv \int \varrho \, d^3 r\,,
\ee
 is guaranteed by the covariant conservation of the modified energy-momentum tensor, despite the fact that the standard energy-momentum tensor is normally not conserved in these theories. Here, 
$\varrho$ is the proper modified energy density inside a particle
\bq
\varrho &=& \mathfrak T_{0 0} = -\frac{2}{ \sqrt{-g}} \frac{\delta(\sqrt{-g}\mathfrak{L}_{\rm m})}{\delta g^{0 0}} = 
g_{00}\mathfrak{L}_{\rm m} -2 \frac{\delta\mathfrak{L}_{\rm m}}{\delta g^{0 0}} \nonumber \\ 
&=& - \mathfrak{L}_{\rm m[on-shell]} \,, \label{MED}
\eq
where, in the last equality, we have taken into account that $g_{00}=-1$ at all points inside the particle in the chosen reference frame and that $\delta\mathfrak{L}_{\rm m}/{\delta g^{0 0}}=0$ for static matter fields (see \cite{Avelino:2018qgt} for further details). Using Eqs. (\ref{MVL}) and (\ref{MED}) it is then  straightforward to show that
\bq
\int \mathfrak T d^3 r &=& \int (-\mathfrak T_{0 0}+ 3 \mathcal P) d^3 r = - \int \mathfrak T_{0 0}d^3 r\nonumber\\
&=&\int  \mathfrak{L}_{\rm m[on-shell]} d^3 r \,.
\eq
Hence, the equality $\mathfrak L_{\rm m[on-shell]}=\mathfrak T$ must hold on average inside a particle in $\mathcal R (R)+\mathcal F(T)$ gravity, again assuming  that the perturbations to the Minkowski metric field play a negligible role on the particle structure. This in turn implies that the equality $\mathcal L_{\rm m[on-shell]} = \mathfrak T - \mathcal F(T)$ must be valid on average inside a particle. Only when $\mathcal F(T)$ plays no significant role in the particle structure will the standard (average) result $\mathcal L_{\rm m[on-shell]} =  T$ hold.

\section{Particles in $1+1$ dimensions \label{SPART}}

In this section we shall consider a simple model allowing for stable localized particles with fixed rest mass and structure in $1+1$ dimensions. We will use it in order to illustrate the breakdown of the standard von Laue condition reported in the previous section and to discuss its implications regarding the form of the matter on-shell Lagrangian.

For simplicity, we shall consider a $f(R,T)$ theory of gravity where
\be
f(R,T)= R+\epsilon T\,,
\ee
and the matter fields are described by the Lagrangian
\be
{\mathcal L}_{\rm m}(\phi,X)= X - V(\phi)\,. 
\ee
with
\be
X=-\frac12 g^{\mu \nu} \nabla_\mu \phi  \nabla_\nu \phi \,.
\ee
Here, $\epsilon$ is a real constant, $\phi$ is a real scalar field, $g_{\mu\nu}$ are the components of the metric tensor and $V(\phi)$ is a scalar field potential. In this case, the energy-momentum tensor is given by
\be
T_{\mu \nu} = \nabla_\mu \phi \nabla_\nu \phi + \mathcal L g_{\mu \nu}\,,
\ee
and its trace is equal to
\be
T= 2 X - 4 V\,.
\ee
For concreteness, we shall assume the following form for the scalar field potential
\be
V(\phi)=\frac{\lambda}{4} (\phi^2 - \eta^2)^2\,,
\ee
where $\lambda$ is a real coupling constant and $\pm \eta$ are the minima of $V(\phi)$.  Notice that the above model is equivalent to general relativity with a modified matter Lagrangian equal to
\be
{\mathfrak L}_{\rm m}={\mathcal L}_{\rm m}+\epsilon T = (1+2\epsilon)(X -{\mathfrak E} V)\,, 
\ee
where ${\mathfrak E}=(1+4\epsilon)/(1+2\epsilon)$. The components of the corresponding modified energy-momentum tensor are given by
\be
{\mathfrak T}_{\mu \nu}=(1+2\epsilon) \nabla_\mu \phi \nabla_\nu \phi + {\mathfrak L}_{\rm m} g_{\mu \nu}\,. \label{MEMT}
\ee

In a $1+1$ dimensional Minkowski space-time with line element $ds^2=-dt^2+dz^2$, the equation of motion for the scalar field $\phi$ is
\be
{\ddot \phi} - \phi''= -{\mathfrak E}\frac{d V}{d \phi}\,, \label{phieqmM}\\
\ee
where a dot denotes a derivative with respect to the physical time $t$ and a prime represents a derivative with respect to the space coordinate $z$.
Consider a static particle (so that $\phi=\phi(z)$) located at $z=0$. In this case Eq. (\ref{phieqmM}) becomes
\be
\phi''= {\mathfrak E}\frac{dV}{d\phi} \label{phieqmM1}\,,
\ee
and it can be integrated to give
\be
\frac{\phi'^2}{2} = {\mathfrak E}V\,,\label{KeqU}
\ee
taking into account that $|\phi| \to \eta$ for $z \to \pm \infty$. Equation (\ref{KeqU}) has the following solution
\be
\phi = \pm \eta \tanh\left(\frac{z}{{\sqrt 2}\delta}\right)\,,
\ee
with
\be
\delta=(\mathfrak E\lambda)^{-1/2} \eta^{-1}\,.
\ee

The components of the energy-momentum tensor can now be written as
\bq
\rho=T_{00}&=&\frac{\phi'^2}{2}+V=(1+\mathfrak E)V\,,\\
T_{0z}&=&T_{z0}=0\,,\\
p=T_{zz}&=&\frac{\phi'^2}{2}-V(\phi)=(\mathfrak E-1)V =\frac{\mathfrak E-1}{1+\mathfrak E} \rho\,.
\eq
Hence, if $\epsilon \neq 0$ (or, equivalently, $\mathfrak E\neq 1$) the proper pressure inside the particle is not zero ---  the equation of state parameter is a constant equal to $w\equiv p/\rho= ({\mathfrak E-1})({1+\mathfrak E}) \neq 0$. This result implies that the particle, of proper mass
\bq
m&=&\int_{- \infty}^{\infty} \rho \, dz = 2 (1+\mathfrak E)\int_{- \infty}^{\infty} V dz \nonumber \\
&=& \frac{8 {\sqrt 2}}{3} (1+\mathfrak E)V_0 \delta  = \frac{2{\sqrt 2}}{3}\frac{1+\mathfrak E}{\mathfrak E^{1/2}}\lambda^{1/2}  \eta^3\,, \label{mass}
\eq
has a non-vanishing average pressure, which represents a breakdown of the standard von Laue condition (in Eq. (\ref{mass}) $V_0 \equiv V(\phi=0) = \lambda \eta^4/4$). The matter on-shell Lagrangian is equal to
\be
{\mathcal L}_{\rm m[on-shell]}=-\left(\frac{\phi'^2}{2}+V\right) = -(1+\mathfrak E)V = -\rho\,,
\ee
with the trace of the energy-momentum tensor being equal to
\be
T=-\rho+p=-\frac{2\rho}{1+ \mathfrak E}\,.
\ee
This implies that if $\epsilon \neq 0$ (or, equivalently, $\mathfrak E\neq 1$)  then ${\mathcal L}_{\rm m[on-shell]} \neq T$.  Notice that ${\mathcal L}_{\rm m[on-shell]} = T$ in general relativity ($\epsilon = 0$).

On the other hand, from Eqs. (\ref{MEMT}) and (\ref{KeqU}), one finds that
\bq
\varrho= {\mathfrak T}_{00}&=&(1+2\epsilon)\left(\frac{\phi'^2}{2}+{\mathfrak E}V\right)=2{\mathfrak E}(1+2\epsilon)V\,,\\
{\mathfrak T}_{0z}&=&{\mathfrak T}_{z0}=0\,,\\
\mathcal P = {\mathfrak T}_{zz}&=&(1+2\epsilon)\left(\frac{\phi'^2}{2}-{\mathfrak E}V\right)=0\,,
\eq
so that the modified matter on-shell Lagrangian satisfies
\be
{\mathfrak L}_{\rm m[on-shell]}=-(1+2\epsilon)\left(\frac{\phi'^2}{2}+{\mathfrak E}V\right) = {\mathfrak T}\,,
\ee
where $\mathfrak T \equiv {\mathfrak T^\mu}_\mu$. 

Although the main purpose of this section was to consider a simple illustrative example in 1+1 dimensions of the breakdown of the standard von Laue condition in  $R+\epsilon T$ gravity, these results also apply to the proper transverse pressure of domain walls in 3+1 dimensions. Unlike in general relativity, the proper transverse pressure of domain walls is in general not zero in $\mathcal R (R) +\mathcal F (T)$ gravity (but the modified transverse pressure must vanish).
	
\section{Particles and their cosmic fluids}

In Secs. \ref{SVLC} and \ref{SPART} we have shown that the energy-momentum tensor is in general not conserved in $f(R,T)$ gravity and that, as a consequence,  the standard von Laue condition normally breaks down in this context. As a result, the volume integral of the proper pressure $p$ inside a localized particle of fixed proper mass and structure is generally nonvanishing. The energy-momentum tensor of a fluid composed of such particles is just the averaged sum of the energy-momentum tensors of the individual particles. This implies that, in $f(R,T)$ gravity, dust --- here defined as a perfect fluid whose particles are at rest in the fluid's proper frame --- is in general not pressureless. Nevertheless, a generalized von Laue condition has been derived in Sec. \ref{SVLC} according to which the volume integral of the proper modified pressure $\mathcal P$ inside a localized particle of fixed proper mass and structure vanishes in $\mathcal R (R)+\mathcal F(T)$ gravity. This in turn allows us to define essential properties of a fluid composed of such particles.

The components of the modified energy-momentum tensor of a perfect fluid composed of many individual particles are given by
\begin{equation}
\label{eq:pfemt}
{\mathfrak T}_{\mu\nu[\rm f]}=(\varrho_{\rm f} + \mathcal P_{\rm f}) \, U_\mu U_\nu + \mathcal P_{\rm f} g_{\mu\nu}\,,
\end{equation}
where $\varrho_{\rm f}$, $\mathcal P_{\rm f}$ and $U_\mu$ are, respectively, the proper modified energy density, the proper modified pressure and the components of the 4-velocity of the fluid. The generalized von Laue condition implies that $\mathcal P_{\rm f}=0$ in the case of dust, but, more generally, that $\mathcal P_{\rm f} = \varrho_{\rm f} \langle v^2 \rangle/3 $, where $\langle v^2 \rangle$ is the mean squared velocity of the particles measured in the proper frame of each fluid element. 

The equivalence with general relativity, when the modified energy-momentum tensor is replaced by the standard one, implies that the dynamical equations for the evolution of a homogeneous and isotropic universe dominated by either relativistic or non-relativistic particles will be identical to those in general relativity, except for the replacement of the proper energy density and pressure of the corresponding fluids by the modified ones.  Hence, the standard cosmological evolution is recovered in the radiation and matter eras in the context of $R+\mathcal F(T)$ gravity. This result is in disagreement with the claim made in \cite{Haghani:2023uad} --- wrongly assuming dust to be pressureless --- that the dynamics of a matter dominated universe in $R + \mathcal F(T)$ gravity is significantly modified with respect to general relativity. It highlights the importance of a careful assessment of common assumptions regarding the properties of the cosmic fluids when considering modified theories of gravity.

In Sec.  \ref{SVLC} we have shown that the condition $\mathcal L_{\rm m} = \mathfrak T - \mathcal F(T)$ is valid on average inside a particle in the context of $\mathcal R (R)+\mathcal F(T)$ gravity. This result also applies to collections of particles whose Lagrangian is just the sum of the Lagrangians of the individual particles and, in particular, to an ideal gas. In this case one has $\mathcal L_{\rm IG[on-shell]} = \mathfrak T_{\rm IG} - \mathcal F(T_{\rm IG})$, where now $\mathcal L_{\rm IG[on-shell]}$ is the on-shell Lagrangian of the ideal gas, $T_{\rm IG}$ is the trace of its energy-momentum tensor, and $\mathfrak T_{\rm IG}$ is the trace of its modified energy-momentum tensor. Only when $\mathcal F(T)$ plays no significant role in the particle structure will the standard  result $\mathcal L_{\rm IG[on-shell]} =  T_{\rm IG}$ for an ideal gas hold.

A key assumption behind our results is that the particles which compose the fluid do not change their mass and structure. Although this condition can be easily accommodated in the context of $\mathcal R (R)+\mathcal F (T)$ gravity, that is not normally the case when more general $f(R,T)$ gravity models are considered. Although we do not explore in detail these more general classes of $f(R,T)$ gravity models in the present work, we expect then to be severely constrained observationally \cite{Planck:2015bue}.

\section{Conclusions}\label{sec:conc}

In this work we have shown that the  standard von-Laue condition does not generally hold in the Minkowski limit of $f(R,T)$ gravity or of other classes of modified gravity theories in which the conservation of energy and momentum is not guaranteed in that limit. As a result, the volume-averaged pressure inside particles is in general nonvanishing in these theories. Nevertheless, we have shown that it is possible to generalize the von Laue condition so that it remains true in the context of $\mathcal R (R)+\mathcal F(T)$ gravity, by considering a covariantly conserved modified energy-momentum tensor --- the generalized von Laue condition may also apply to more general $f(R,T)$ models as long as the covariant conservation of the modified energy-momentum tensor is approximately valid. We also determined the  impact of our findings on the form of the on-shell Lagrangian of an ideal gas in this context. 

We have discussed the implications of our results for the properties of a perfect fluid composed of many individual particles, showing that standard  results do hold in $\mathcal R (R)+\mathcal F(T)$ gravity, independently of the specific properties of $\mathcal F(T)$, as long as its proper pressure, proper energy density and on-shell lagrangian are replaced by their modified versions. We have further shown that our results have profound cosmological implications. In particular, they imply that, contrary to recent claims, the evolution of a universe dominated by relativistic or non-relativistic particles in $R + \mathcal F(T)$ gravity is the same as in general relativity (in the case of $\mathcal R (R)+ \mathcal F(T)$ gravity the evolution of such a universe would also be independent of $\mathcal F(T)$). As a consequence, the standard radiation and matter era evolution is recovered in $R + \mathcal F(T)$ gravity. 

These results can be of fundamental importance for cosmological studies involving $f(R,T)$ gravity, or other theories of gravity in which the equations of motion for the matter fields differ from those in general relativity in the Minkowski limit. 
	
\begin{acknowledgments}
	
We thank Lara Sousa, Rui Azevedo, Samuel Veiga, Vasco Ferreira, and my other colleagues of the Cosmology group at Instituto de Astrofíısica e Ciências do Espaço for enlightening discussions. We acknowledge the support by Fundação para a Ciência e a Tecnologia (FCT) through the research Grants No. UIDB/04434/2020 and No. UIDP/04434/2020. This work was also supported by FCT through the R$\&$D project 2022.03495.PTDC - {\it Uncovering the nature of cosmic strings}.	

\end{acknowledgments}
 
\bibliography{fRT1_PLB}

\begin{thebibliography}{36}
\expandafter\ifx\csname natexlab\endcsname\relax\def\natexlab#1{#1}\fi
\expandafter\ifx\csname bibnamefont\endcsname\relax
  \def\bibnamefont#1{#1}\fi
\expandafter\ifx\csname bibfnamefont\endcsname\relax
  \def\bibfnamefont#1{#1}\fi
\expandafter\ifx\csname citenamefont\endcsname\relax
  \def\citenamefont#1{#1}\fi
\expandafter\ifx\csname url\endcsname\relax
  \def\url#1{\texttt{#1}}\fi
\expandafter\ifx\csname urlprefix\endcsname\relax\def\urlprefix{URL }\fi
\providecommand{\bibinfo}[2]{#2}
\providecommand{\eprint}[2][]{\url{#2}}

\bibitem[{\citenamefont{Laue}(1911)}]{doi:10.1002/andp.19113400808}
\bibinfo{author}{\bibfnamefont{M.}~\bibnamefont{Laue}},
  \bibinfo{journal}{Annalen der Physik} \textbf{\bibinfo{volume}{340}},
  \bibinfo{pages}{524} (\bibinfo{year}{1911}).

\bibitem[{\citenamefont{Avelino and Sousa}(2018)}]{Avelino:2018qgt}
\bibinfo{author}{\bibfnamefont{P.~P.} \bibnamefont{Avelino}} \bibnamefont{and}
  \bibinfo{author}{\bibfnamefont{L.}~\bibnamefont{Sousa}},
  \bibinfo{journal}{Phys. Rev.} \textbf{\bibinfo{volume}{D97}},
  \bibinfo{pages}{064019} (\bibinfo{year}{2018}), \eprint{1802.03961}.

\bibitem[{\citenamefont{Avelino}(2023)}]{Avelino:2023rac}
\bibinfo{author}{\bibfnamefont{P.~P.} \bibnamefont{Avelino}},
  \bibinfo{journal}{JCAP} \textbf{\bibinfo{volume}{08}}, \bibinfo{pages}{005}
  (\bibinfo{year}{2023}), \eprint{2303.06630}.

\bibitem[{\citenamefont{Batool et~al.}(2024)\citenamefont{Batool, Sultan, Olmo,
  and Rubiera-Garcia}}]{Batool:2024axw}
\bibinfo{author}{\bibfnamefont{A.}~\bibnamefont{Batool}},
  \bibinfo{author}{\bibfnamefont{A.~M.} \bibnamefont{Sultan}},
  \bibinfo{author}{\bibfnamefont{G.~J.} \bibnamefont{Olmo}}, \bibnamefont{and}
  \bibinfo{author}{\bibfnamefont{D.}~\bibnamefont{Rubiera-Garcia}},
  \bibinfo{journal}{Phys. Rev. D} \textbf{\bibinfo{volume}{110}},
  \bibinfo{pages}{064059} (\bibinfo{year}{2024}), \eprint{2407.06062}.

\bibitem[{\citenamefont{Schutz}(1970)}]{Schutz:1970my}
\bibinfo{author}{\bibfnamefont{B.~F.} \bibnamefont{Schutz}},
  \bibinfo{journal}{Phys. Rev. D} \textbf{\bibinfo{volume}{2}},
  \bibinfo{pages}{2762} (\bibinfo{year}{1970}).

\bibitem[{\citenamefont{Schutz and Sorkin}(1977)}]{Schutz:1977df}
\bibinfo{author}{\bibfnamefont{B.~F.} \bibnamefont{Schutz}} \bibnamefont{and}
  \bibinfo{author}{\bibfnamefont{R.}~\bibnamefont{Sorkin}},
  \bibinfo{journal}{Annals Phys.} \textbf{\bibinfo{volume}{107}},
  \bibinfo{pages}{1} (\bibinfo{year}{1977}).

\bibitem[{\citenamefont{Brown}(1993)}]{Brown:1992kc}
\bibinfo{author}{\bibfnamefont{J.~D.} \bibnamefont{Brown}},
  \bibinfo{journal}{Class. Quant. Grav.} \textbf{\bibinfo{volume}{10}},
  \bibinfo{pages}{1579} (\bibinfo{year}{1993}), \eprint{gr-qc/9304026}.

\bibitem[{\citenamefont{Faraoni}(2009)}]{Faraoni:2009rk}
\bibinfo{author}{\bibfnamefont{V.}~\bibnamefont{Faraoni}},
  \bibinfo{journal}{Phys. Rev. D} \textbf{\bibinfo{volume}{80}},
  \bibinfo{pages}{124040} (\bibinfo{year}{2009}), \eprint{0912.1249}.

\bibitem[{\citenamefont{Capozziello et~al.}(2018)\citenamefont{Capozziello,
  Mantica, and Molinari}}]{Capozziello:2018ddp}
\bibinfo{author}{\bibfnamefont{S.}~\bibnamefont{Capozziello}},
  \bibinfo{author}{\bibfnamefont{C.~A.} \bibnamefont{Mantica}},
  \bibnamefont{and} \bibinfo{author}{\bibfnamefont{L.~G.}
  \bibnamefont{Molinari}}, \bibinfo{journal}{Int. J. Geom. Meth. Mod. Phys.}
  \textbf{\bibinfo{volume}{16}}, \bibinfo{pages}{1950008}
  (\bibinfo{year}{2018}), \eprint{1810.03204}.

\bibitem[{\citenamefont{Ferreira et~al.}(2020)\citenamefont{Ferreira, Avelino,
  and Azevedo}}]{Ferreira:2020fma}
\bibinfo{author}{\bibfnamefont{V.~M.~C.} \bibnamefont{Ferreira}},
  \bibinfo{author}{\bibfnamefont{P.~P.} \bibnamefont{Avelino}},
  \bibnamefont{and} \bibinfo{author}{\bibfnamefont{R.~P.~L.}
  \bibnamefont{Azevedo}}, \bibinfo{journal}{Phys. Rev. D}
  \textbf{\bibinfo{volume}{102}}, \bibinfo{pages}{063525}
  (\bibinfo{year}{2020}), \eprint{2005.07739}.

\bibitem[{\citenamefont{Capozziello and
  De~Laurentis}(2011)}]{Capozziello:2011et}
\bibinfo{author}{\bibfnamefont{S.}~\bibnamefont{Capozziello}} \bibnamefont{and}
  \bibinfo{author}{\bibfnamefont{M.}~\bibnamefont{De~Laurentis}},
  \bibinfo{journal}{Phys. Rept.} \textbf{\bibinfo{volume}{509}},
  \bibinfo{pages}{167} (\bibinfo{year}{2011}), \eprint{1108.6266}.

\bibitem[{\citenamefont{Olmo}(2011)}]{Olmo:2011uz}
\bibinfo{author}{\bibfnamefont{G.~J.} \bibnamefont{Olmo}},
  \bibinfo{journal}{Int. J. Mod. Phys. D} \textbf{\bibinfo{volume}{20}},
  \bibinfo{pages}{413} (\bibinfo{year}{2011}), \eprint{1101.3864}.

\bibitem[{\citenamefont{Clifton et~al.}(2012)\citenamefont{Clifton, Ferreira,
  Padilla, and Skordis}}]{Clifton:2011jh}
\bibinfo{author}{\bibfnamefont{T.}~\bibnamefont{Clifton}},
  \bibinfo{author}{\bibfnamefont{P.~G.} \bibnamefont{Ferreira}},
  \bibinfo{author}{\bibfnamefont{A.}~\bibnamefont{Padilla}}, \bibnamefont{and}
  \bibinfo{author}{\bibfnamefont{C.}~\bibnamefont{Skordis}},
  \bibinfo{journal}{Phys. Rept.} \textbf{\bibinfo{volume}{513}},
  \bibinfo{pages}{1} (\bibinfo{year}{2012}), \eprint{1106.2476}.

\bibitem[{\citenamefont{Berti et~al.}(2015)}]{Berti:2015itd}
\bibinfo{author}{\bibfnamefont{E.}~\bibnamefont{Berti}} \bibnamefont{et~al.},
  \bibinfo{journal}{Class. Quant. Grav.} \textbf{\bibinfo{volume}{32}},
  \bibinfo{pages}{243001} (\bibinfo{year}{2015}), \eprint{1501.07274}.

\bibitem[{\citenamefont{Nojiri et~al.}(2017)\citenamefont{Nojiri, Odintsov, and
  Oikonomou}}]{Nojiri:2017ncd}
\bibinfo{author}{\bibfnamefont{S.}~\bibnamefont{Nojiri}},
  \bibinfo{author}{\bibfnamefont{S.~D.} \bibnamefont{Odintsov}},
  \bibnamefont{and} \bibinfo{author}{\bibfnamefont{V.~K.}
  \bibnamefont{Oikonomou}}, \bibinfo{journal}{Phys. Rept.}
  \textbf{\bibinfo{volume}{692}}, \bibinfo{pages}{1} (\bibinfo{year}{2017}),
  \eprint{1705.11098}.

\bibitem[{\citenamefont{Odintsov et~al.}(2023)\citenamefont{Odintsov,
  Oikonomou, Giannakoudi, Fronimos, and Lymperiadou}}]{Odintsov:2023weg}
\bibinfo{author}{\bibfnamefont{S.~D.} \bibnamefont{Odintsov}},
  \bibinfo{author}{\bibfnamefont{V.~K.} \bibnamefont{Oikonomou}},
  \bibinfo{author}{\bibfnamefont{I.}~\bibnamefont{Giannakoudi}},
  \bibinfo{author}{\bibfnamefont{F.~P.} \bibnamefont{Fronimos}},
  \bibnamefont{and} \bibinfo{author}{\bibfnamefont{E.~C.}
  \bibnamefont{Lymperiadou}}, \bibinfo{journal}{Symmetry}
  \textbf{\bibinfo{volume}{15}}, \bibinfo{pages}{1701} (\bibinfo{year}{2023}),
  \eprint{2307.16308}.

\bibitem[{\citenamefont{Nesseris}(2009)}]{Nesseris:2008mq}
\bibinfo{author}{\bibfnamefont{S.}~\bibnamefont{Nesseris}},
  \bibinfo{journal}{Phys. Rev. D} \textbf{\bibinfo{volume}{79}},
  \bibinfo{pages}{044015} (\bibinfo{year}{2009}), \eprint{0811.4292}.

\bibitem[{\citenamefont{Harko and Lobo}(2010)}]{Harko:2010mv}
\bibinfo{author}{\bibfnamefont{T.}~\bibnamefont{Harko}} \bibnamefont{and}
  \bibinfo{author}{\bibfnamefont{F.~S.~N.} \bibnamefont{Lobo}},
  \bibinfo{journal}{Eur. Phys. J. C} \textbf{\bibinfo{volume}{70}},
  \bibinfo{pages}{373} (\bibinfo{year}{2010}), \eprint{1008.4193}.

\bibitem[{\citenamefont{Harko et~al.}(2011)\citenamefont{Harko, Lobo, Nojiri,
  and Odintsov}}]{Harko:2011kv}
\bibinfo{author}{\bibfnamefont{T.}~\bibnamefont{Harko}},
  \bibinfo{author}{\bibfnamefont{F.~S.~N.} \bibnamefont{Lobo}},
  \bibinfo{author}{\bibfnamefont{S.}~\bibnamefont{Nojiri}}, \bibnamefont{and}
  \bibinfo{author}{\bibfnamefont{S.~D.} \bibnamefont{Odintsov}},
  \bibinfo{journal}{Phys. Rev. D} \textbf{\bibinfo{volume}{84}},
  \bibinfo{pages}{024020} (\bibinfo{year}{2011}), \eprint{1104.2669}.

\bibitem[{\citenamefont{Kat\i{}rc\i{} and Kavuk}(2014)}]{Katirci:2013okf}
\bibinfo{author}{\bibfnamefont{N.}~\bibnamefont{Kat\i{}rc\i{}}}
  \bibnamefont{and} \bibinfo{author}{\bibfnamefont{M.}~\bibnamefont{Kavuk}},
  \bibinfo{journal}{Eur. Phys. J. Plus} \textbf{\bibinfo{volume}{129}},
  \bibinfo{pages}{163} (\bibinfo{year}{2014}), \eprint{1302.4300}.

\bibitem[{\citenamefont{Haghani et~al.}(2013)\citenamefont{Haghani, Harko,
  Lobo, Sepangi, and Shahidi}}]{Haghani:2013oma}
\bibinfo{author}{\bibfnamefont{Z.}~\bibnamefont{Haghani}},
  \bibinfo{author}{\bibfnamefont{T.}~\bibnamefont{Harko}},
  \bibinfo{author}{\bibfnamefont{F.~S.~N.} \bibnamefont{Lobo}},
  \bibinfo{author}{\bibfnamefont{H.~R.} \bibnamefont{Sepangi}},
  \bibnamefont{and} \bibinfo{author}{\bibfnamefont{S.}~\bibnamefont{Shahidi}},
  \bibinfo{journal}{Phys. Rev. D} \textbf{\bibinfo{volume}{88}},
  \bibinfo{pages}{044023} (\bibinfo{year}{2013}), \eprint{1304.5957}.

\bibitem[{\citenamefont{Ludwig et~al.}(2015)\citenamefont{Ludwig, Minazzoli,
  and Capozziello}}]{Ludwig:2015hta}
\bibinfo{author}{\bibfnamefont{H.}~\bibnamefont{Ludwig}},
  \bibinfo{author}{\bibfnamefont{O.}~\bibnamefont{Minazzoli}},
  \bibnamefont{and}
  \bibinfo{author}{\bibfnamefont{S.}~\bibnamefont{Capozziello}},
  \bibinfo{journal}{Phys. Lett. B} \textbf{\bibinfo{volume}{751}},
  \bibinfo{pages}{576} (\bibinfo{year}{2015}), \eprint{1506.03278}.

\bibitem[{\citenamefont{Harko et~al.}(2018)\citenamefont{Harko, Koivisto, Lobo,
  Olmo, and Rubiera-Garcia}}]{Harko:2018gxr}
\bibinfo{author}{\bibfnamefont{T.}~\bibnamefont{Harko}},
  \bibinfo{author}{\bibfnamefont{T.~S.} \bibnamefont{Koivisto}},
  \bibinfo{author}{\bibfnamefont{F.~S.~N.} \bibnamefont{Lobo}},
  \bibinfo{author}{\bibfnamefont{G.~J.} \bibnamefont{Olmo}}, \bibnamefont{and}
  \bibinfo{author}{\bibfnamefont{D.}~\bibnamefont{Rubiera-Garcia}},
  \bibinfo{journal}{Phys. Rev. D} \textbf{\bibinfo{volume}{98}},
  \bibinfo{pages}{084043} (\bibinfo{year}{2018}), \eprint{1806.10437}.

\bibitem[{\citenamefont{Bahamonde}(2018)}]{Bahamonde:2017ifa}
\bibinfo{author}{\bibfnamefont{S.}~\bibnamefont{Bahamonde}},
  \bibinfo{journal}{Eur. Phys. J. C} \textbf{\bibinfo{volume}{78}},
  \bibinfo{pages}{326} (\bibinfo{year}{2018}), \eprint{1709.05319}.

\bibitem[{\citenamefont{Barrientos et~al.}(2018)\citenamefont{Barrientos, Lobo,
  Mendoza, Olmo, and Rubiera-Garcia}}]{Barrientos:2018cnx}
\bibinfo{author}{\bibfnamefont{E.}~\bibnamefont{Barrientos}},
  \bibinfo{author}{\bibfnamefont{F.~S.~N.} \bibnamefont{Lobo}},
  \bibinfo{author}{\bibfnamefont{S.}~\bibnamefont{Mendoza}},
  \bibinfo{author}{\bibfnamefont{G.~J.} \bibnamefont{Olmo}}, \bibnamefont{and}
  \bibinfo{author}{\bibfnamefont{D.}~\bibnamefont{Rubiera-Garcia}},
  \bibinfo{journal}{Phys. Rev. D} \textbf{\bibinfo{volume}{97}},
  \bibinfo{pages}{104041} (\bibinfo{year}{2018}), \eprint{1803.05525}.

\bibitem[{\citenamefont{Minazzoli}(2018)}]{Minazzoli:2018xjy}
\bibinfo{author}{\bibfnamefont{O.}~\bibnamefont{Minazzoli}},
  \bibinfo{journal}{Phys. Rev. D} \textbf{\bibinfo{volume}{98}},
  \bibinfo{pages}{124020} (\bibinfo{year}{2018}), \eprint{1811.05845}.

\bibitem[{\citenamefont{Fox}(2019)}]{Fox:2018gop}
\bibinfo{author}{\bibfnamefont{M.~S.} \bibnamefont{Fox}},
  \bibinfo{journal}{Phys. Rev. D} \textbf{\bibinfo{volume}{99}},
  \bibinfo{pages}{124027} (\bibinfo{year}{2019}), \eprint{1810.11595}.

\bibitem[{\citenamefont{Asimakis et~al.}(2023)\citenamefont{Asimakis,
  Basilakos, Lymperis, Petronikolou, and Saridakis}}]{Asimakis:2022jel}
\bibinfo{author}{\bibfnamefont{P.}~\bibnamefont{Asimakis}},
  \bibinfo{author}{\bibfnamefont{S.}~\bibnamefont{Basilakos}},
  \bibinfo{author}{\bibfnamefont{A.}~\bibnamefont{Lymperis}},
  \bibinfo{author}{\bibfnamefont{M.}~\bibnamefont{Petronikolou}},
  \bibnamefont{and} \bibinfo{author}{\bibfnamefont{E.~N.}
  \bibnamefont{Saridakis}}, \bibinfo{journal}{Phys. Rev. D}
  \textbf{\bibinfo{volume}{107}}, \bibinfo{pages}{104006}
  (\bibinfo{year}{2023}), \eprint{2212.03821}.

\bibitem[{\citenamefont{Gon\c{c}alves et~al.}(2024)\citenamefont{Gon\c{c}alves,
  Rosa, and Lobo}}]{Goncalves:2023umv}
\bibinfo{author}{\bibfnamefont{T.~B.} \bibnamefont{Gon\c{c}alves}},
  \bibinfo{author}{\bibfnamefont{J.~a.~L.} \bibnamefont{Rosa}},
  \bibnamefont{and} \bibinfo{author}{\bibfnamefont{F.~S.~N.}
  \bibnamefont{Lobo}}, \bibinfo{journal}{Phys. Rev. D}
  \textbf{\bibinfo{volume}{109}}, \bibinfo{pages}{084008}
  (\bibinfo{year}{2024}), \eprint{2305.05337}.

\bibitem[{\citenamefont{Avelino and Azevedo}(2018)}]{Avelino:2018rsb}
\bibinfo{author}{\bibfnamefont{P.~P.} \bibnamefont{Avelino}} \bibnamefont{and}
  \bibinfo{author}{\bibfnamefont{R.~P.~L.} \bibnamefont{Azevedo}},
  \bibinfo{journal}{Phys. Rev. D} \textbf{\bibinfo{volume}{97}},
  \bibinfo{pages}{064018} (\bibinfo{year}{2018}), \eprint{1802.04760}.

\bibitem[{\citenamefont{Avelino and Azevedo}(2022)}]{Avelino:2022eqm}
\bibinfo{author}{\bibfnamefont{P.~P.} \bibnamefont{Avelino}} \bibnamefont{and}
  \bibinfo{author}{\bibfnamefont{R.~P.~L.} \bibnamefont{Azevedo}},
  \bibinfo{journal}{Phys. Rev. D} \textbf{\bibinfo{volume}{105}},
  \bibinfo{pages}{104005} (\bibinfo{year}{2022}), \eprint{2203.04022}.

\bibitem[{\citenamefont{Haghani et~al.}(2024)\citenamefont{Haghani, Harko, and
  Shahidi}}]{Haghani:2023uad}
\bibinfo{author}{\bibfnamefont{Z.}~\bibnamefont{Haghani}},
  \bibinfo{author}{\bibfnamefont{T.}~\bibnamefont{Harko}}, \bibnamefont{and}
  \bibinfo{author}{\bibfnamefont{S.}~\bibnamefont{Shahidi}},
  \bibinfo{journal}{Phys. Dark Univ.} \textbf{\bibinfo{volume}{44}},
  \bibinfo{pages}{101448} (\bibinfo{year}{2024}), \eprint{2301.12133}.

\bibitem[{\citenamefont{Avelino}(2024)}]{Avelino:2024rub}
\bibinfo{author}{\bibfnamefont{P.~P.} \bibnamefont{Avelino}},
  \bibinfo{journal}{Phys. Rev. D} \textbf{\bibinfo{volume}{110}},
  \bibinfo{pages}{024064} (\bibinfo{year}{2024}), \eprint{2404.12373}.

\bibitem[{\citenamefont{Fisher and Carlson}(2019)}]{Fisher:2019ekh}
\bibinfo{author}{\bibfnamefont{S.~B.} \bibnamefont{Fisher}} \bibnamefont{and}
  \bibinfo{author}{\bibfnamefont{E.~D.} \bibnamefont{Carlson}},
  \bibinfo{journal}{Phys. Rev. D} \textbf{\bibinfo{volume}{100}},
  \bibinfo{pages}{064059} (\bibinfo{year}{2019}), \eprint{1908.05306}.

\bibitem[{\citenamefont{Akarsu et~al.}(2024)\citenamefont{Akarsu,
  Bouhmadi-L\'opez, Kat\i{}rc\i{}, Nazari, Roshan, and Uzun}}]{Akarsu:2023lre}
\bibinfo{author}{\bibfnamefont{O.}~\bibnamefont{Akarsu}},
  \bibinfo{author}{\bibfnamefont{M.}~\bibnamefont{Bouhmadi-L\'opez}},
  \bibinfo{author}{\bibfnamefont{N.}~\bibnamefont{Kat\i{}rc\i{}}},
  \bibinfo{author}{\bibfnamefont{E.}~\bibnamefont{Nazari}},
  \bibinfo{author}{\bibfnamefont{M.}~\bibnamefont{Roshan}}, \bibnamefont{and}
  \bibinfo{author}{\bibfnamefont{N.~M.} \bibnamefont{Uzun}},
  \bibinfo{journal}{Phys. Rev. D} \textbf{\bibinfo{volume}{109}},
  \bibinfo{pages}{104055} (\bibinfo{year}{2024}), \eprint{2306.11717}.

\bibitem[{\citenamefont{Ade et~al.}(2016)}]{Planck:2015bue}
\bibinfo{author}{\bibfnamefont{P.~A.~R.} \bibnamefont{Ade}}
  \bibnamefont{et~al.} (\bibinfo{collaboration}{Planck}),
  \bibinfo{journal}{Astron. Astrophys.} \textbf{\bibinfo{volume}{594}},
  \bibinfo{pages}{A14} (\bibinfo{year}{2016}), \eprint{1502.01590}.

\end{thebibliography}
 	
 \end{document}